\documentclass[a4paper]{jpconf}
\usepackage{graphicx}
\newcommand{\ba}{\begin{eqnarray}}
\newcommand{\ea}{\end{eqnarray}}
\newcommand{\ban}{\begin{eqnarray*}}
\newcommand{\ean}{\end{eqnarray*}}
\newcommand{\bsub}{\begin{subequations}}
\newcommand{\esub}{\end{subequations}}
\newcommand{\tl}{\tilde{\ell}}
\newcommand{\tlam}{\tilde{\Lambda}}
\newcommand{\nc}{\newcommand}
\nc{\Id}{{\mathchoice {\rm 1\mskip-4mu l} {\rm 1\mskip-4mu l}
{\rm 1\mskip-4.5mu l} {\rm 1\mskip-5mu l}}}
\begin{document}
\title{Deformed pseudospin doublets as a fingerprint of\\ 
a relativistic supersymmetry in nuclei}

\author{A. Leviatan}

\address{Racah Institute of Physics, The Hebrew University, 
Jerusalem 91904, Israel}

\ead{ami@phys.huji.ac.il}

\begin{abstract}
The single-particle spectrum of deformed shell-model states 
in nuclei, is shown to exhibit a supersymmetric pattern. The latter 
involves deformed pseudospin doublets and intruder levels. 
The underlying supersymmetry is associated with the relativistic 
pseudospin symmetry of the nuclear mean-field Dirac Hamiltonian 
with scalar and vector potentials.
\end{abstract}

\section{Introduction}
The concept of pseudospin doublets~\cite{hecht69,arima69} is based on 
the empirical observation in nuclei of quasi-degenerate pairs of 
certain normal-parity shell-model orbitals with non-relativistic 
single-nucleon quantum numbers 
\ba
(n, \ell, j = \ell+1/2)
\quad {\rm and} \quad
(n-1, \ell+2, j = \ell+3/2) ~.
\label{psdoubsp}
\ea 
The doublet structure is expressed in terms of a ``pseudo'' orbital 
angular momentum, $\tilde{\ell} = \ell+1$, and ``pseudo'' spin, 
$\tilde {s} = 1/2$,
which are coupled to $j = \tilde{\ell} \pm \tilde {s}$. 
For example, $(1d_{5/2},0g_{7/2})$ and $(2s_{1/2},1d_{3/2})$ 
will have $\tl= 3$ and $\tl =1$, respectively.
The states $(n=0,\ell,j=\ell+1/2)$, 
with aligned spin and no nodes, are not part of a doublet. 
For large $j$ they are the ``intruder'' abnormal-parity states, 
{\it i.e.}, $0g_{9/2},\;0h_{11/2},\;0i_{13/2}$, 
which are unique in the major shell. 
In the presence of axially-symmetric deformation, the doublets 
persist~\cite{bohr82} with asymptotic (Nilsson) quantum numbers 
\ba
[N,n_3,\Lambda]\Omega=\Lambda+1/2
\quad {\rm and} \quad
[N,n_3,\Lambda +2]\Omega=\Lambda+3/2 ~.
\label{psdoub}
\ea
Here $N$, $n_3$, are harmonic oscillator quantum numbers, 
$\Lambda$ and $\Omega$ are, respectively, the components of the orbital 
and total angular momentum along the symmetry (3rd) axis.   
In this case, the doublets can be expressed in terms of pseudo-orbital, 
$\tilde{\Lambda}=\Lambda+1$, and pseudospin, $\tilde{\mu}=\pm 1/2$, 
projections, to obtain $\Omega=\tilde{\Lambda} \pm 1/2$. 
For example, $([420]1/2,[422]3/2)$ and $([411]3/2,[413]5/2)$ 
will have $\tilde{\Lambda}=1$ and $\tilde{\Lambda}=2$, respectively. 
The intruder nodeless states, $[N,n_3,\Lambda=N-n_3]\Omega=\Lambda+1/2$, 
are not part of a doublet. 
Pseudospin doublets play a central role in explaining 
features of nuclei~\cite{bohr82}, including 
superdeformation~\cite{dudek87} and identical bands~\cite{naza90,step90}. 
They have been shown to originate from a relativistic symmetry of the 
Dirac Hamiltonian in which the sum of the scalar and vector mean-field 
potentials cancel~\cite{gino97}. 
The generators of this relativistic SU(2) symmetry have been 
identified~\cite{ginolev98} and consequences for radial nodes 
have been considered~\cite{levgino01}. 
Recently, it has been shown that a pseudospin-invariant Dirac Hamiltonian 
with spherical potentials, gives rise to a 
supersymmetric pattern~\cite{lev04}. 
In the present contribution we extend this study and show that 
a supersymmetric pattern occurs also when the relevant 
potentials are axially deformed~\cite{lev09}. 

\begin{figure}[t]
\begin{minipage}{20pc}
\includegraphics[width=9.6pc,angle=270]{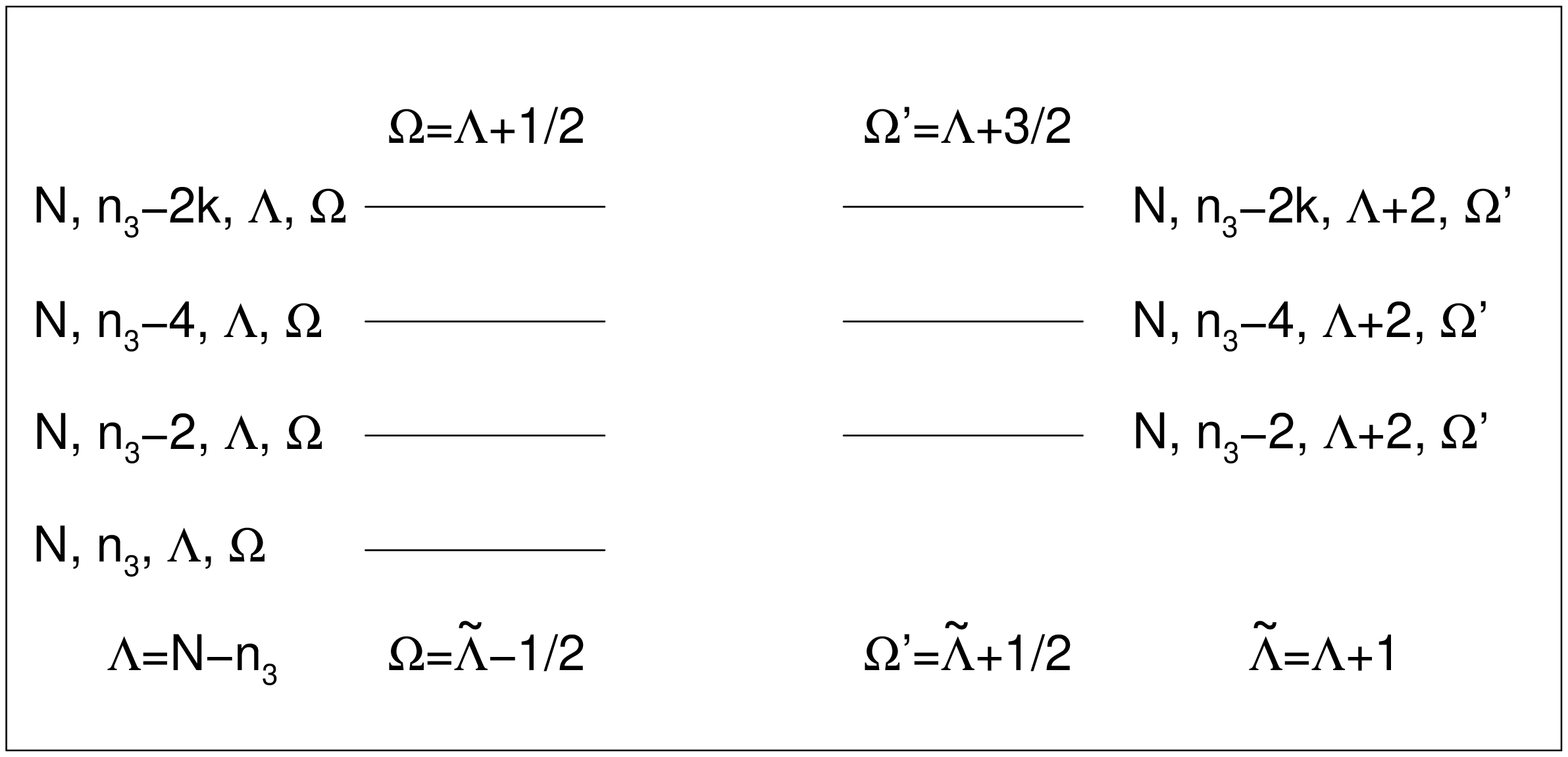}
\caption{\label{fig1}
Nuclear single-particle spectrum composed of deformed pseudospin 
doublets and an intruder level. 
All states share a common $\tilde{\Lambda}$.}
\end{minipage}\hspace{3pc}%
\begin{minipage}{14pc}
\includegraphics[width=8.5pc,angle=270]{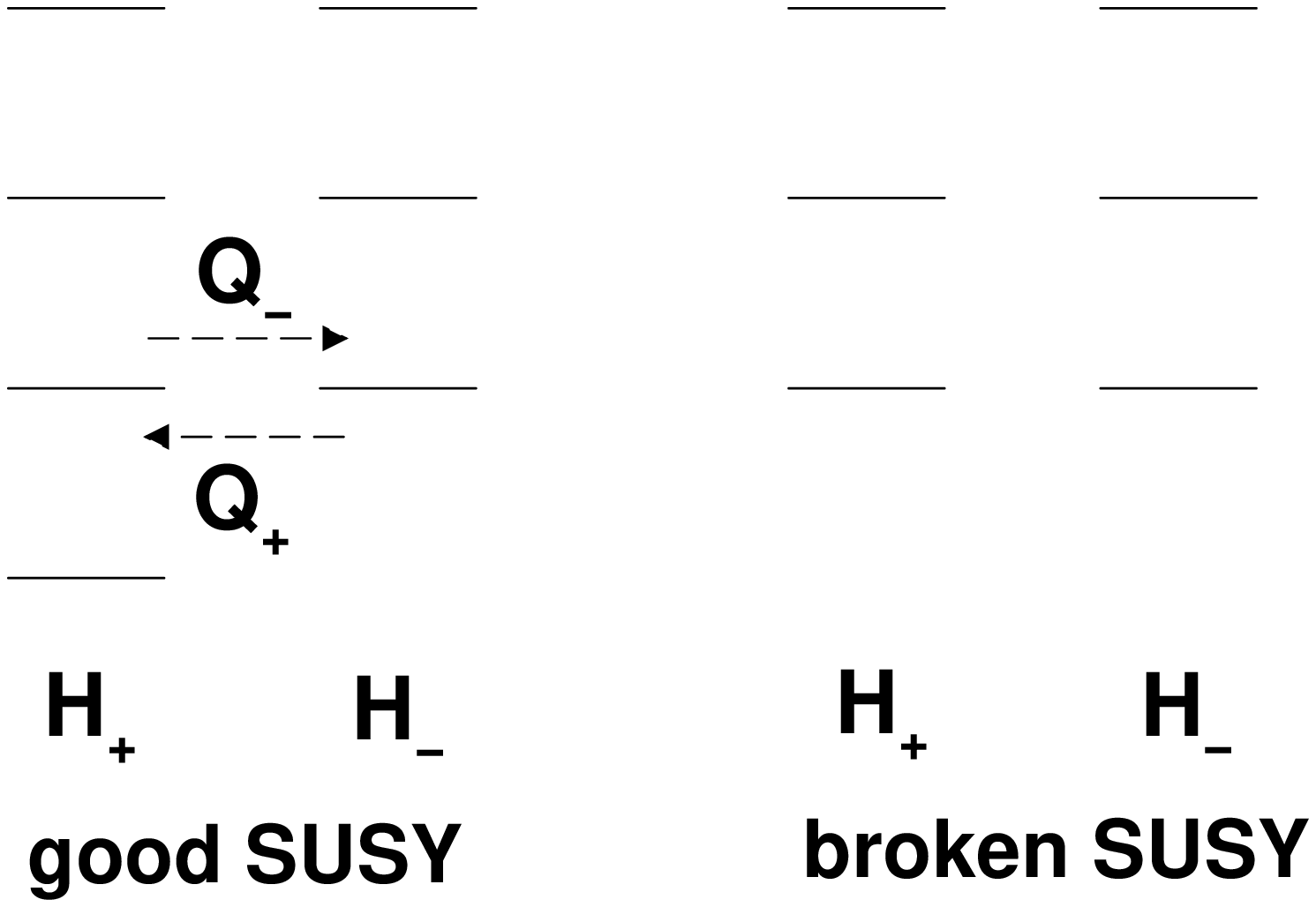}
\caption{\label{fig2}
Typical spectra of good and broken SUSY. The operators $Q_{-}$ 
and $Q_{+}$ connect degenerate states in the $H_{+}$ and $H_{-}$ 
sectors.}
\end{minipage} 
\end{figure}

Fig.~\ref{fig1} portrays the level scheme of a subset of 
deformed pseudospin doublets, Eq.~(\ref{psdoub}), with fixed 
$N$, $\Lambda$, $\Omega$, $\Omega^{\prime}$ 
and $n_{\perp}= (N-\Lambda -n_3)/2=1,2,3,\ldots$ 
together with the intruder level 
$[N,n_3,\Lambda=N-n_3]\Omega=\Lambda+1/2$, 
(or equivalently, $n_{\perp}=0$). 
The single-particle spectrum exhibits 
towers of pair-wise degenerate states, 
sharing a common $\tilde{\Lambda}$, and an additional non-degenerate 
nodeless intruder state at the bottom of the 
spin-projection aligned tower. 
A~comparison with Fig.~2 reveals a striking similarity with 
a supersymmetric pattern. 
In what follows we identify 
the underlying supersymmetric structure associated with a 
Dirac Hamiltonian possessing a relativistic pseudospin symmetry. 

\section{Dirac Hamiltonian with axially-deformed potentials}

A relativistic mean-field description of nuclei employs a 
Dirac Hamiltonian, 
$H = \mbox{\boldmath $\hat{\alpha}\cdot \hat{p}$}
+ \hat{\beta} (M  + V_S) + V_V$ for a nucleon of mass~$M$ 
moving in external scalar, $V_S$, and vector,
$V_V$, potentials. 
When the potentials are axially-symmetric, {\it i.e.}, independent 
of the azimuthal angle~$\phi$, $V_{S,V}=V_{S,V}(\rho,z)\,$, 
$\rho = \sqrt{x^2+y^2}$, then 
the $z$-component of the angular momentum 
operator, $\hat{J}_z$, commutes with 
$H$ and its half-integer eigenvalues $\Omega$ are used to label the 
Dirac wave functions 
\ba
\Psi_{\Omega}(\rho,\phi,z) = 
\left ( 
\begin{array}{c}
g^{+}(\rho,z)\, e^{i(\Omega - 1/2)\phi}\\
g^{-}(\rho,z)\, e^{i(\Omega + 1/2)\phi}\\
if^{+}(\rho,z)\, e^{i(\Omega - 1/2)\phi}\\
if^{-}(\rho,z)\, e^{i(\Omega + 1/2)\phi}\\
\end{array}
\right ) ~.
\label{wf1}
\ea
Here $g^{\pm}\equiv g^{\pm}(\rho,z)$ and $
f^{\pm}\equiv f^{\pm}(\rho,z)$ are the radial wave functions of the 
upper and lower components, respectively. 
The Dirac equation, $H\Psi = E\Psi$, leads to a set of four coupled 
partial differential equations for these components. 
For each solution with $\Omega>0$, 
there is a degenerate time-reversed solution 
with $-\Omega <0$, hence, 
we confine the discussion to solutions with $\Omega>0$. 
Of particular interest 
are bound Dirac valence states with 
$ 0 < E < M$ and normalizable wave functions.  
The potentials enter the Dirac equation through the combinations 
\ba
\label{A}
A(\rho,z) &=& E + M + V_S(\rho,z) - V_V(\rho,z) ~,
\nonumber\\
\label{B}
B(\rho,z) &=& E - M - V_S(\rho,z) - V_V(\rho,z) ~.
\label{AB}
\ea
For relativistic mean-fields relevant to nuclei, 
$V_S$ is attractive and $V_V$ is repulsive with typical values 
$V_S(0) \sim -400,\; V_V(0)\sim 350,$ MeV. 
The potentials satisfy 
$\rho V_S(\rho,z),\rho V_V(\rho,z) \rightarrow 0$ 
for $\rho\rightarrow 0$ and 
$V_S(\rho,z),V_V(\rho,z) \rightarrow 0$ 
for $\rho\rightarrow \infty$ or $z\rightarrow\pm\infty$. 
The boundary conditions imply that the radial wave functions 
fall off exponentially for large distances and behave as 
a power law for $\rho\rightarrow 0$. 
Furthermore, for $z=0$ and $\rho\rightarrow \infty$, 
$f^{-}/g^{+} \propto (M-E)>0$ and $g^{-}/f^{+} \propto (M+E) > 0$, 
while for $z=0$ and $\rho\rightarrow 0$, 
$f^{-}/g^{+}\propto B(0)\rho$ and 
$g^{-}/f^{+}\propto -A(0)\rho$.
These properties have important implications for the 
structure of radial nodes. In particular, it follows that 
for potentials with the indicated asymptotic behaviour and 
$A(0),\,B(0)>0$, as encountered in nuclei, 
a necessary condition for a nodeless bound 
eigenstate of a Dirac Hamiltonian is~\cite{lev09} 
\ba
g^{-} =0 \;\; {\rm or}\;\; f^{+} =0 ~.
\label{nodeless}
\ea

\section{Relativistic pseudospin symmetry in nuclei}

A relativistic pseudospin symmetry occurs when 
the sum of the scalar and vector potentials is a constant~\cite{gino97} 
\ba
V_{S}(\rho,z) + V_{V}(\rho,z) = \Delta_0 ~. 
\label{D0}
\ea
The symmetry generators, ${\hat{\tilde {S}}}_{i}$, 
commute with the Dirac Hamiltonian 
and span an SU(2) algebra~\cite{ginolev98}
\ba
{\hat{\tilde {S}}}_{i} = 
\left (
\begin{array}{cc}
 U_p\, \hat{s}_i\, U_p &  0 \\
0 & \hat{s}_{i}
\end{array}
\right ) 
\quad\;\; i =x,y,z 
\quad\;\;\; U_p = \, \frac{\mbox{\boldmath $\sigma\cdot p$}}{p} ~.
\label{pSgen}
\ea
Here 
${\hat s}_{i} = \sigma_{i}/2$ are the usual spin operators, 
defined in terms of Pauli matrices. 
The Dirac eigenfunctions in the pseudospin limit satisfy
\ba
{\hat{\tilde {S}}}_{z}
\Psi^{(\tilde{\mu})}_{\Omega} &=&
\tilde{\mu}\Psi^{(\tilde{\mu})}_{\Omega}\qquad
\;\tilde{\mu} = \pm 1/2
\nonumber\\ 
({\hat{\tilde {S}}}_{x} + i{\hat{\tilde {S}}}_{y})
\Psi^{(-1/2)}_{\Omega} &=&
\Psi^{(1/2)}_{\Omega}
\ea
and form degenerate $SU(2)$ doublets. 
Their wave functions have been shown to be of 
the form~\cite{gino02}
\ba
\Psi^{(-1/2)}_{\Omega_1=\tilde{\Lambda}-1/2} 
= 
\left ( 
\begin{array}{c}
g^{+}\, e^{i(\tilde{\Lambda} - 1)\phi}\\
-g\, e^{i\tilde{\Lambda}\phi}\\
0 \\
if\, e^{i\tilde{\Lambda}\phi}\\
\end{array}
\right )
\quad , \quad
\Psi^{(1/2)}_{\Omega_2=\tilde{\Lambda}+1/2} 
= 
\left ( 
\begin{array}{c}
g\, e^{i\tilde{\Lambda}\phi}\\
g^{-}\, e^{i(\tilde{\Lambda}+1)\phi}\\
if\, e^{i\tilde{\Lambda}\phi}\\
0 \\
\end{array}
\right )
\label{wfps}
\ea
where $\tilde{\Lambda} = \Omega-\tilde{\mu}\geq 0$ 
is the eigenvalue of $\hat{J}_z - {\hat{\tilde {S}}}_{z}$.
Near the pseudospin limit, eigenstates with $\tilde{\Lambda}\neq 0$ 
occur in degenerate doublets. 
An exception to this rule are eigenstates whose radial 
components have no nodes and, therefore, are not part of a doublet. 
The latter property follows from the fact that 
a nodeless bound Dirac state satisfies the criteria 
of Eq.~(\ref{nodeless}), hence has a wave function 
$\Psi^{(-1/2)}_{\Omega_1=\tilde{\Lambda}-1/2}$, as in 
Eq.~(\ref{wfps}), with $g^{+},\, g,\, f\neq 0$ and 
$f/g^{+} >0$. Its pseudospin partner state has a wave function 
$\Psi^{(1/2)}_{\Omega_2=\tilde{\Lambda}+1/2}$. 
The radial components satisfy 
$Bg^{-} = [B - 2(\tilde{\Lambda}/\rho)f/g^{+}]g^{+}$, 
where $B$ is defined in Eq.~(\ref{B}). 
This relation is satisfied, to a good approximation, 
for mean-field potentials relevant to nuclei, and the r.h.s. 
is non-zero and, consequently, $g^{-}\neq 0$. If so, then 
the partner state, $\Psi^{(1/2)}_{\Omega_2=\tilde{\Lambda}+1/2}$, 
is also nodeless, but it cannot be a bound eigenstate 
since its radial components do not fulfill 
the condition of Eq.~(\ref{nodeless}). 

The relativistic pseudospin symmetry has 
been tested in numerous realistic mean-field calculations 
of nuclei and was found to be obeyed to a good approximation, 
especially for doublets near the Fermi 
surface~\cite{gino02,ginomad98,lala98,meng99,ginolev01,ginolev04,gino05}. 
The dominant upper components of the states in Eq.~(\ref{wfps}),  
involving $g^{+}$ and $g^{-}$, correspond to the non-relativistic 
pseudospin doublets of Eq.~(\ref{psdoub}). 
The nodeless Dirac eigenstates correspond to the non-relativistic 
intruder states, $[N,n_3,\Lambda=N-n_3]\Omega=\Lambda+1/2$, mentioned above. 
In what follows we show that the ensemble of deformed pseudospin doublets 
and intruder levels, forms a supersymmetric pattern whose 
origin can be traced to the relativistic pseudospin 
symmetry of the Dirac Hamiltonian.

\section{Relativistic pseudospin symmetry and 
supersymmetric quantum mechanics}

The essential ingredients of supersymmetric quantum 
mechanics~\cite{junker96} 
are the supersymmetric Hamiltonian, ${\cal H}$,  
and charges $Q_{+}$, $Q_{-} = Q_{+}^{\dagger}$, 
which generate the supersymmetry (SUSY) algebra
\ba
\left [\,{\cal H}\,,\,Q_{\pm}\,\right] = 0 \;\;\; , \;\;\; 
\left \{\,Q_{\pm}\, , \, Q_{\pm}\,\right\} = 0 \;\;\; , \;\;\; 
\left\{\,Q_{-}\, ,\, Q_{+}\,\right\}= {\cal H} ~.
\label{susyalg}
\ea
Accompanying this set is a Hermitian $Z_2$-grading 
operator satisfying 
\ba
{\cal P}^{\dagger} = {\cal P}\;\;\; , \;\;\;
\left [\, {\cal H}\, , \, {\cal P}\,\right ] =0\;\;\; , \;\;\; 
\left \{\, Q_{\pm}\, , \, {\cal P}\,\right \} =0\;\;\; , \;\;\;
{\cal P}^2 = \Id ~.
\label{grading}
\ea 
The $+1$ and $-1$ eigenspaces of ${\cal P}$ define the ``positive-parity'', 
$H_{+}$, and ``negative-parity'', $H_{-}$, sectors of the spectrum, with 
eigenvectors $\Psi^{(+)}$ and $\Psi^{(-)}$, respectively. 
The SUSY algebra imply that if $\Psi^{(+)}$ is 
an eigenstate of ${\cal H}$, then also 
$\Psi^{(-)}=Q_{-}\Psi^{(+)}$ is an 
eigenstate of ${\cal H}$ with the same energy, 
unless $Q_{-}\Psi^{(+)}$ vanishes or produces an unphysical state, 
({\it e.g.}, non-normalizable). 
The resulting spectrum consists of pairwise degenerate levels 
with a non-degenerate single state (the ground state) 
in one sector when the supersymmetry 
is exact. If all states are pairwise degenerate, 
the supersymmetry is said to be broken. 
Typical spectra for good and broken SUSY are shown in Fig.~\ref{fig2}.
\begin{figure}[t]
\includegraphics[width=\linewidth]{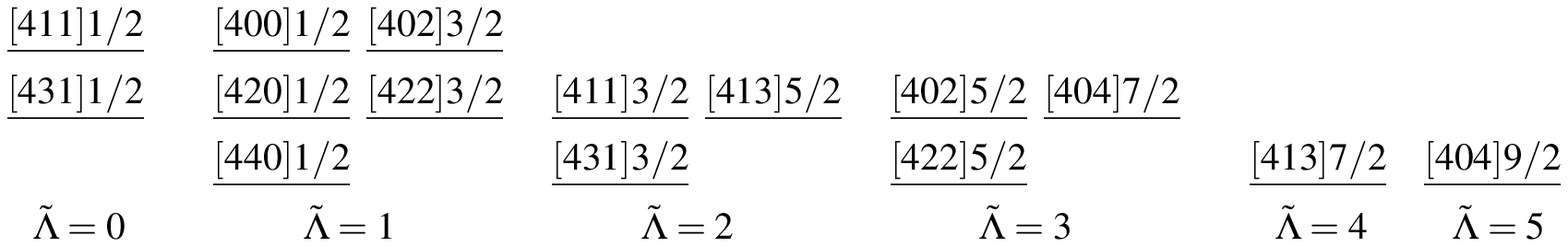}
\caption{\label{fig3}
Grouping of deformed shell-model states
$[N=4,n_3,\Lambda]\Omega$, exhibiting 
a pattern of good SUSY, relevant 
to the pseudospin symmetry limit~\cite{lev09}.}
\end{figure}

Degenerate doublets, signaling a supersymmetric structure, 
can emerge in a quantum system with a Hamiltonian $H$, from the existence 
of two Hermitian, conserved and anticommuting operators, $\hat{R}$ and 
$\hat{B}$
\ba
[ H, \hat{R}] = [H, \hat{B}] = \{\hat{R},\hat{B}\} =0~.  
\label{HRB}
\ea
The operator $\hat{R}$ has non-zero eigenvalues, $r$, 
which come in pairs of opposite signs. 
$\hat{B}^2 = \hat{B}^{\dagger}\hat{B} = f(H)$, is a function 
of the Hamiltonian. A $Z_2$-grading operator, 
${\cal P}_r $, and Hermitian supersymmetric charges, $Q_1$, $Q_2$ 
and Hamiltonian ${\cal H}$ can now be constructed as 
\ba
{\cal P}_r = \hat{R}/\vert r \vert \;\;\; , \;\;\; 
Q_1 = \hat{B}\;\;\; , \;\;\; Q_2 = iQ_1{\cal P}_{r} \;\;\; , \;\;\;
{\cal H} = Q_{1}^2=f(H) ~.
\ea
The triad of operators 
$Q_{\pm} = (Q_1 \pm iQ_2)/2$ and ${\cal H}$ 
form the standard SUSY algebra of Eq.~(\ref{susyalg}). 

Focusing the discussion to the relativistic pseudospin limit, 
the elements of the 
SUSY algebra are 
\ba 
&&Q_1 = 2\left (\, M +\Delta_0 - H\,\right )
{\hat{\tilde {S}}}_{x}\;\; , \;\;
Q_2 = 2\left (\, M +\Delta_0 - H\,\right )
{\hat{\tilde {S}}}_{y}
\;\; , \;\;
{\cal P}_r = 2{\hat{\tilde {S}}}_{z}\;\; , \;\;
\nonumber\\
&&{\cal H} = Q_{1}^2 = \left (\, M +\Delta_0 - H\,\right )^2 ~,
\ea 
where $H$ is the pseudospin-invariant Dirac Hamiltonian with potentials 
satisfying the condition of Eq.~(\ref{D0}). 
The operator $\hat{B} = Q_1$ 
connects the doublet states of Eq.~(\ref{wfps}). 
The spectrum, for each $\tilde{\Lambda}\neq 0$, 
consists of twin towers of pairwise degenerate pseudospin doublet states,  
with $\Omega_1=\tilde{\Lambda}-1/2$ and $\Omega_2=\tilde{\Lambda}+1/2$, 
and an additional non-degenerate 
nodeless state at the bottom of the $\Omega_1=\tilde{\Lambda}-1/2$ tower. 
Altogether, the ensemble of Dirac states with $\Omega_2-\Omega_1=1$ exhibits 
a supersymmetric pattern of good SUSY, as illustrated 
in Fig.~\ref{fig3}.

The supersymmetry discussed above, 
is one of several possible classes of supersymmetries of 
Dirac Hamiltonians in 3+1 dimensions, 
with cylindrically--deformed scalar and vector potentials~\cite{lev09}. 
The supersymmetries associated with the pseudospin and spin limits 
arise when the potentials obey a constraint on their sum or 
difference, respectively. 
Additional supersymmetries arise when one of 
the potentials has a $1/\rho$ dependence and the second 
potential depends on $z$. The relevant $\hat{R}$ and $\hat{B}$ 
operators (\ref{HRB}) of the various supersymmetries, 
are listed in Table~\ref{Tab1}. A~similar analysis can be done for 
spherical scalar and vector potentials and, as shown 
in Table~\ref{Tab2}, the corresponding supersymmetries are associated 
with the pseudospin ($V_S+V_V=\Delta_0)$, 
spin ($V_S-V_V=\Xi_0$) and Coulomb ($V_S,V_V\propto 1/r$) 
limits of the Dirac Hamiltonian~\cite{lev04,levgin10}. 
\begin{center}
\begin{table}[t]
\caption{\label{Tab1}
Conserved, anticommuting operators, $\hat{R}$ and $\hat{B}$,  
for Dirac Hamiltonians $(H)$, with axially-deformed scalar and vector 
potentials, exhibiting a supersymmetric 
structure~\cite{lev09}. Here ${\hat{\tilde {S}}}_{i} = 
\left ({U_p\, \hat{s}_i\, U_p\atop 0}{0\atop \hat{s}_{i}}\right )$ 
and ${\hat{S}}_{i} = \left ({\hat{s}_{i}\atop 0}
{0\atop U_p\,\hat{s}_i\, U_p}\right )$
are the pseudospin and spin generators, respectively. 
$\hat{R}_{z} = \left [\,M+V_{S}(z)\,\right ]\hat{\beta}\,\hat{\Sigma}_3 
+\gamma_{5}\hat{p}_z$ and  
$\hat{R}_{\rho} =
[\,M+V_{S}(\rho)\,]\hat{\Sigma}_3 
-i\hat{\beta}\,\gamma_{5}
(\,\mbox{\boldmath $\hat{\Sigma}$}\times
\mbox{\boldmath $\hat{p}$}\,)_{3}$, 
where 
$\hat{\Sigma}_i = \left ({\sigma_i\atop 0}{0\atop \sigma_i}\right )$.}
\smallskip
\centering
\begin{tabular}{lllll}
\br
Dirac Hamiltonian & $\hat{R}$ 
& \hspace{1.1cm} $\hat{B}$ & \hspace{0.7cm}$\hat{B}^2=f(H)$ & SUSY \\
\mr
$V_{S}(\rho,z) + V_{V}(\rho,z) = \Delta_0\;$
& $\hat{\tilde {S}}_{z}$ & 
 $2(M+\Delta_0 - H)\hat{\tilde {S}}_{x}$ & 
 $( M+\Delta_0 - H)^2$ & good \\[2pt] 
$V_{S}(\rho,z) - V_{V}(\rho,z) = \Xi_0\;$
& $\hat{S}_{z}$ &
 $ 2(\,M+\Xi_0 + H)\,\hat{S}_{x}$ &
 $ (M+\Xi_0 + H)^2 $ & broken \\[2pt]
$V_S = V_{S}(z)$, $V_V = \frac{\alpha_{V}}{\rho}$ 
& $\hat{R}_{z}\; $ & 
$\hat{\beta}\,\hat{\Sigma}_3
\{i\hat{J}_{z}\gamma_5 [\,H -  \hat{\Sigma}_3\hat{R}_z ]\;$ & 
$\hat{J}_{z}^2 ( H^2 - \hat{R}_{z}^2) 
+ \alpha_{V}^2 \hat{R}_{z}^2$ & good\\[2pt]
 & &
$\qquad\,\, -\frac{\alpha_{V}}{\rho}
(\mbox{\boldmath $\hat{\Sigma}\cdot\rho$})\,\hat{R}_z \} \;$ & \\[2pt] 
$V_S = \frac{\alpha_{S}}{\rho}$, $V_V = V_{V}(z)$ 
& $\hat{R}_{\rho}\;$  & 
$\hat{\Sigma}_3\{i\hat{J}_{z}\gamma_5 [M - \hat{\Sigma}_3 
\hat{R}_{\rho}] \;$ & 
$\hat{J}_{z}^2 
(\hat{R}_{\rho}^2 - M^2) 
+ \alpha_{S}^2 \hat{R}_{\rho}^2$ & good \\[2pt]
 & & 
$\qquad\,
-\frac{\alpha_{S}}{\rho} 
(\mbox{\boldmath $\hat{\Sigma}\cdot\rho$}) 
\hat{\beta}\hat{R}_{\rho}\}$ & \\[2pt] 
\br
\end{tabular}
\end{table}
\end{center}
\begin{center}
\begin{table}[t]
\caption{\label{Tab2} 
As in Table~\ref{Tab1} but for spherical scalar and vector 
potentials~\cite{lev04,levgin10} in the Dirac Hamiltonian $(H)$. 
Here
$\hat{K} = 
-\hat{\beta}
\left (\, \mbox{\boldmath $\hat{\Sigma}\cdot\hat{\ell}$} +1\right )$, 
with $\mbox{\boldmath $\hat{{\ell}}$} = \mbox{\boldmath ${r}$}\times 
\mbox{\boldmath $\hat{p}$}$, 
$\hat{\tilde T}_z = 
(\mbox{\boldmath $\hat{\tilde {S}}$}\times 
\mbox{\boldmath $
\hat{\tilde {L}}$})_z$, 
$\hat{T}_z = 
(\mbox{\boldmath $\hat{S}$}\times 
\mbox{\boldmath $
\hat{L}$})_z$. 
$\hat{\tilde{Y}} = \mbox{\boldmath $\hat{\tilde{L}}\cdot\hat{\tilde{L}}$}
+1/4 - \hat{J}^{2}_{z} $ and 
$\hat{Y}= \mbox{\boldmath $\hat{L}\cdot\hat{L}$}
+1/4 - \hat{J}^{2}_{z} $. The operators 
$\hat{\Sigma}_i$, ${\hat{\tilde {S}}}_{i}$ and ${\hat{S}}_{i}$ are 
defined in the caption of Table~\ref{Tab1} and 
$\hat{\tilde{L}} = \hat{J}_i - {\hat{\tilde {S}}}_{i}$, 
$\hat{L} = \hat{J}_i - \hat{S}_{i}$.}
\smallskip
\centering
\begin{tabular}{lllll}
\br 
Dirac Hamiltonian & \hspace{0.4cm} $\hat{R}$ 
& \hspace{1.1cm} $\hat{B}$ & \hspace{0.5cm} $\hat{B}^2=f(H)$ & SUSY\\
\mr
$V_{S}(r) + V_{V}(r) = \Delta_0\;$
& $\hat{K}-1/2$ &
$2(\,M+\Delta_0 - H )\,\hat{\tilde{T}}_z$ & 
$( M+\Delta_0 - H )^2\,\hat{\tilde{Y}} \;$ & good \\[2pt] 
$V_{S}(r) - V_{V}(r) = \Xi_0\quad$
& $\hat{K}+1/2$ &
$2(\,M+\Xi_0 + H )\,\hat{T}_z$ &
$( M+\Xi_0 + H )^2\,\hat{Y}\;$ & broken\\[2pt] 
$V_S = \frac{\alpha_S}{r}$, $V_V = \frac{\alpha_{V}}{r}$ 
& $\hat{K}$ & 
$ -i\hat{K}\gamma_5\,(H - \hat{\beta}M )\qquad\;\;\;\;$ & 
$(\alpha_{V}M + \alpha_{S}H )^2\;$ & good\\ 
& & 
$\;\; + 
\frac{1}{r}(\mbox{\boldmath $\hat{\Sigma}\cdot r$}) 
(\alpha_{V} M + \alpha_{S} H )$ & 
$\;\;\; + \hat{K}^2 ( H^2 - M^2 )$ & \\[1pt]
\br
\end{tabular}
\end{table}
\end{center}

\section{Summary}

As previously noted, the single-particle spectrum of spherical 
shell-model states in nuclei, exhibits characteristic features of 
a supersymmetric pattern. The latter consists of twin towers 
of pair-wise degenerate spherical pseudospin doublets 
(with $n=1,2\ldots$) sharing a common $\tilde{\ell}$, and an additional 
non-degenerate nodeless ($n=0$) intruder state at the bottom of the 
spin-aligned tower
\ba
\begin{array}{ccl}
\left (n,\, \ell,\, j = \ell + 1/2 \,\right )\quad  &
\left (n -1,\, \ell + 2,\, j = \ell + 3/2 \,\right )\quad & 
n = 1,2,\ldots \\
\left (n=0,\, \ell,\, j = \ell + 1/2 \,\right ) \quad & & n =0 \\
 & & \\ 
j = \tl - 1/2 &  
j = \tl + 1/2 & \,\tl = \ell + 1 \\ 
\end{array}
\ea
As discussed in the present contribution, a similar situation is observed 
in the single-particle spectrum of deformed shell model states. 
The corresponding supersymmetric pattern consists of twin towers of 
pair-wise degenerate deformed pseudospin doublets 
(with $n_{\perp}=1,2\ldots$) sharing a common $\tilde{\Lambda}$, 
and an additional non-degenerate nodeless ($n_{\perp}=0$) intruder state 
at the bottom of the spin-projection aligned tower 
\ba
\begin{array}{ccl}
\left [ N,n_3 - 2n_{\perp},\Lambda\right ]\Omega = \Lambda +1/2
\;\;\;  & 
\left [ N,n_3 - 2n_{\perp},\Lambda+ 2\right ]\Omega = \Lambda +1/2
\;\;\;  & 
n_{\perp} = 1,2,\ldots \\
\left [ N,n_3,\Lambda\right ]\Omega = \Lambda +1/2\;\;  
& &  n_{\perp} = 0 \\
& & \\ 
\Omega = \tlam - 1/2 &  
\Omega = \tlam + 1/2 & \;\;\tlam = \Lambda + 1 \\
\end{array}
\ea
with $\Lambda=N-n_3$. 
In both cases, the underlying supersymmetric algebra is associated 
with the relativistic pseudospin symmetry of the nuclear mean-field 
Dirac Hamiltonian, involving scalar and vector potentials. 
Additional supersymmetries of Dirac Hamiltonians with such mixed Lorentz 
structure can be identified. 

\section*{Acknowledgments}
This work is supported by the Israel Science Foundation.


\section*{References}
\begin{thebibliography}{9}

\bibitem{hecht69}
Hecht K T and A. Adler A 1969
{\it Nucl. Phys.} A {\bf 137} 129.

\bibitem{arima69}
Arima A, Harvey M and Shimizu K 1969
{\it Phys. Lett.} B {\bf 30} 517.

\bibitem{bohr82}
Bohr A, Hamamoto I and Mottelson B R 1982 
{\it Phys. Scr.} {\bf 26} 267.

\bibitem{dudek87}
Dudek J, Nazarewicz W, Szymanski Z and Leander G A 1987 
{\it Phys. Rev. Lett.} {\bf 59} 1405.

\bibitem{naza90}
Nazarewicz W, Twin P J, Fallon P and Garrett J D 1990 
{\it Phys. Rev. Lett.} {\bf 64} 1654. 

\bibitem{step90}
Stephens F S  {\it et al.} 1990
{\it Phys. Rev. Lett.} {\bf 65} 301.  

\bibitem{gino97}
Ginocchio J N 1997 
{\it Phys. Rev. Lett.} {\bf 78} 436. 

\bibitem{ginolev98}
Ginocchio J N and  Leviatan A 1998 
{\it Phys. Lett.} B {\bf 425} 1.

\bibitem{levgino01}
Leviatan A and Ginocchio J N 2001 
{\it Phys. Lett.} B {\bf 518} 214.

\bibitem{lev04}
Leviatan A 2004 
{\it Phys. Rev. Lett.} {\bf 92} 202501.

\bibitem{lev09} 
Leviatan A 2009
{\it Phys. Rev. Lett.} {\bf 103} 042502.

\bibitem{gino02}
Ginocchio J N 2002 
{\it Phys. Rev. C} {\bf 66} 064312.

\bibitem{ginomad98}
Ginocchio J N and Madland D G 1998 
{\it Phys. Rev. C} {\bf 57} 1167.

\bibitem{lala98}
Lalazissis G A, Gambhir Y K, Maharana J P, Warke C S and Ring P 1998
{\it Phys. Rev. C} {\bf 58} R45. 

\bibitem{meng99}
Meng J, Sugawara-Tanabe K, Yamaji S and Arima A 1999  
{\it Phys. Rev. C} {\bf 59} 154. 

\bibitem{ginolev01}
Ginocchio J N and Leviatan A 2001
{\it Phys. Rev. Lett.} {\bf 87} 072502.

\bibitem{ginolev04}
Ginocchio J N, Leviatan A, Meng J and Zhou S G 2004
{\it Phys. Rev.} C {\bf 69} 034303.

\bibitem{gino05}
Ginocchio J N 2005 
{\it Phys. Rep.} {\bf 414} 165.

\bibitem{junker96}
Junker G 1996 
{\it Supersymmetric Methods in Quantum and Statistical Physics}, 
(Berlin: Springer).

\bibitem{levgin10}
Leviatan A and Ginocchio J N 2010 in preparation.

\end{thebibliography}
\end{document}